\documentclass[12pt, a4paper]{article}
\usepackage{amsmath}

\begin{document}

\title{On  integrability of (2+1)-dimensional quasilinear systems}
\author{E.V. Ferapontov and K.R. Khusnutdinova
  \footnote{On the leave from: Institute of Mechanics, Ufa Branch of 
the Russian Academy of
  Sciences, Karl Marx Str. 6, Ufa, 450000, Russia.}}
  \date{}
  \maketitle
  \vspace{-7mm}
\begin{center}
Department of Mathematical Sciences \\ Loughborough University \\
Loughborough, Leicestershire LE11 3TU \\ United Kingdom \\[2ex]
e-mails: \\[1ex] \texttt{E.V.Ferapontov@lboro.ac.uk}\\
\texttt{K.Khusnutdinova@lboro.ac.uk}
\end{center}

\bigskip

\begin{abstract}
A (2+1)-dimensional  quasilinear system  is said to be `integrable' 
if it can be decoupled in infinitely many ways into a pair of 
compatible $n$-component one-dimensional systems in Riemann 
invariants. Exact solutions described by these reductions, known  as 
nonlinear interactions of planar simple waves, can be viewed as 
natural dispersionless analogs of $n$-gap solutions. It is 
demonstrated that the requirement of the existence of 'sufficiently 
many' $n$-component reductions provides the effective classification 
criterion. As an example of this approach we classify
  integrable (2+1)-dimensional systems of conservation laws possessing 
a convex quadratic entropy.


\end{abstract}

\newpage

\section{Introduction}

In this paper we address the problem of integrability of 
$(2+1)$-dimensional quasilinear systems
\begin{equation}
{\bf u}_t+A({\bf u}) {\bf u}_x+B({\bf u}) {\bf u}_y=0
\label{1}
\end{equation}
where $t, x, y$ are independent variables,  ${\bf u}$ is  an 
$m$-component column vector and $A({\bf u}), B({\bf u})$ are $m\times 
m$ matrices. Systems of this type describe many physical phenomena. 
In particular, important examples occur in gas dynamics, shallow 
water theory, combustion theory, nonlinear elasticity, magneto-fluid 
dynamics, etc. \cite{Majda}. Although
many interesting systems of the form (\ref{1}) arise as 
dispersionless limits of multidimensional
  soliton equations \cite{Zakharov} or within the R-matrix approach 
\cite{Bla} and, therefore,  should be regarded as integrable, no 
intrinsic definition of  the integrability for multidimensional 
quasilinear systems  has been proposed until recently.
In particular, the standard symmetry approach \cite{Mik1, Mik2}, 
which proves to be extremely effective in the case of higher order 
evolution equations and systems,  does not seem to work in this 
context.

The key element of our construction are  exact solutions of the 
system (\ref{1}) of the form
${\bf u}={\bf u}(R^1, ..., R^n)$ where the {\it Riemann invariants} 
$R^1, ..., R^n$ solve a pair of
commuting diagonal systems
\begin{equation}
R^i_t=\lambda^i(R)\ R^i_x, ~~~~ R^i_y=\mu^i(R)\ R^i_x;
\label{R}
\end{equation}
notice that the number of Riemann invariants is allowed to be 
arbitrary!  Thus, the original  $2+1$-dimensional system (\ref{1}) is 
decoupled into a pair of diagonal $1+1$-dimensional systems. 
Solutions of this type, known as nonlinear interactions of $n$ planar 
simple waves, were extensively investigated in gas dynamics and 
magnetohydrodynamics in a series of publications \cite{Burnat1, 
Burnat2, Burnat3, Perad1, Perad2, Dinu, Grundland}. Later, they 
appeared in the context of the dispersionless KP hierarchy 
\cite{Gibb81, Gibb94, GibTsa96, GibTsa99, GuMaAl, MaAl, Ma, Lei} and 
the theory of  integrable hydrodynamic-type chains \cite{Pavlov, 
Pavlov1}. We will call a
multidimensional system {\it integrable} if it possesses 
`sufficiently many' $n$-component reductions of the form (\ref{R}) 
for arbitrary $n$ (the precise definition follows). The corresponding 
nonlinear $n$-wave interactions can be viewed as   dispersionless 
analogs of  `$n$-gap' solutions.

We recall, see \cite{Tsarev}, that the requirement of  the 
commutativity of the flows (\ref{R})
is equivalent to the following restrictions on their characteristic speeds:
\begin{equation}
\frac{\partial_j\lambda 
^i}{\lambda^j-\lambda^i}=\frac{\partial_j\mu^i}{\mu^j-\mu^i}, ~~~ 
i\ne j,
~~~ \partial_j=\partial/\partial_{ R^j};
\label{comm}
\end{equation}
(no summation!). Once these conditions are met, the general solution 
of (\ref{R}) is given by the
implicit  `generalized hodograph'  formula \cite{Tsarev}
\begin{equation}
v^i(R)=x+\lambda^i(R)\ t+\mu^i(R) \ y, ~~~ i=1, ..., n,
\label{hod}
\end{equation}
where $v^i(R)$ are  characteristic speeds of the general flow 
commuting with (\ref{R}), that is, the general solution of the linear 
system
\begin{equation}
\frac{\partial_jv^i}{v^j-v^i}=\frac{\partial_j\lambda 
^i}{\lambda^j-\lambda^i}=\frac{\partial_j\mu^i}{\mu^j-\mu^i}, ~~~ 
i\ne j.
\label{comm1}
\end{equation}
Substituting ${\bf u}(R^1, ..., R^n)$ into (\ref{1}) and using 
(\ref{R}), one readily arrives at the equations
\begin{equation}
(A+\mu^iB+\lambda^i I_m)\ \partial_i{\bf u}=0, ~~~~~ i=1, ..., n,
\label{2}
\end{equation}
implying that both $\lambda^i$ and $\mu^i$ satisfy the dispersion relation
\begin{equation}
{\rm det} (A+\mu B+\lambda I_m)=0.
\label{dispersion}
\end{equation}
Thus, the construction of nonlinear interactions of $n$ planar simple 
waves consists of two steps:

\noindent {\bf (1)} Reduce the initial system (\ref{1}) to a pair of 
commuting flows (\ref{R}) by solving the equations  (\ref{comm}), 
(\ref{2}) for ${\bf u}(R), \ \lambda^i(R), \ \mu^i(R)$ as functions 
of the Riemann invariants $R^1, ..., R^n$. For $n\geq 3$ these 
equations are highly overdetermined and
do not possess solutions in general. However, once a particular 
reduction of the form (\ref{R}) is constructed, the second step is 
fairly straightforward:

\noindent {\bf (2)} Solve the linear system (\ref{comm1}) for 
$v^i(R)$ and determine $R^1, ..., R^n$ as functions of $t, x, y$ from 
the implicit  hodograph formula (\ref{hod}).

{\bf Remark 1.}
For $n=1$ we have
${\bf u}={\bf u}(R)$,  where the scalar variable $R=R^1$ solves a 
pair of first order PDEs
$$
R_t=\lambda (R)\ R_x, ~~~~ R_y=\mu (R)\ R_x
$$
which, in  one-component situation, are automatically commuting. We 
recall that in the scalar case the  hodograph formula (\ref{hod}) 
takes the  form
\begin{equation}
f(R)=x+\lambda(R)t+\mu(R)y
\label{P1}
\end{equation}
where $f(R)$ is  arbitrary. This formula  shows that, in coordinates 
$t, x, y$,  the surfaces $R=const$ are planes. Hence, 
${\bf{u}}={\bf u}(R)$ is constant along a one-parameter family of 
planes. Solutions of this type, known as planar simple waves, exist 
for all multidimensional quasilinear systems  and, therefore, cannot 
detect the integrability.

Similarly, for  $n=2$, we have
${\bf u}={\bf u}(R^1, R^2)$  where
$R^1, R^2$ satisfy the system (\ref{R}) whose general solution is 
given by the generalized hodograph formula
\begin{equation}
v^1(R)=x+\lambda^1(R)t+\mu^1(R)y, ~~~ 
v^2(R)=x+\lambda^2(R)t+\mu^2(R)y, ~~~ R=(R^1, R^2).
\label{P2}
\end{equation}
Setting $R^1=const, ~ R^2=const$, one obtains a two-parameter family 
of lines (or, in the geometric language, a line congruence) in the 
3-space of
independent variables $t, x, y$. Therefore, the solution ${\bf 
u}={\bf u}(R^1, R^2)$ is constant along the lines of a two-parameter 
family. The requirement
of  the existence of solutions of this type, known as nonlinear 
interactions of two planar simple waves, is also not very 
restrictive. For instance, for $m=2$, any  system (\ref{1}) possesses 
infinitely many 2-component reductions of the form (\ref{R}) 
parametrized by two arbitrary functions of a single  argument (see 
examples below).

\medskip

On the contrary, the requirement of  the existence of nontrivial 
3-component reductions is
already sufficiently  restrictive and, in particular, implies the 
existence of  $n$-component reductions for arbitrary $n$. This 
phenomenon is similar to the well-known three-soliton condition in 
the Hirota bilinear approach \cite{Hietarinta1, Hietarinta2, 
Hietarinta3}
(recall that two-soliton solutions exist for arbitrary PDEs 
transformable to Hirota's bilinear form and, therefore, cannot detect 
the integrability),
and the condition of three-dimensional consistency in
the classification of discrete integrable systems on quad-graphs \cite{Adler}.
One can show, by analyzing  equations  (\ref{comm}), (\ref{2}), that 
the maximum number of $n$-component reductions the system (\ref{1}) 
may possess is parametrized, modulo changes of variables $R^i\to 
f^i(R^i)$,  by $n$ arbitrary functions of a single argument (notice 
that this number does not depend on $m$). Therefore, we propose the 
following

{\bf Definition.} {\it A $(2+1)$-dimensional quasilinear system  is 
said to be integrable  if it possesses $n$-component reductions of 
the form (\ref{R})  parametrized  by $n$ arbitrary functions of a 
single argument.}

{\bf Remark 2.} This definition of  integrability implies that the 
matrices $A$ and $B$ in  (\ref{1}) are {\it not} commuting (if we 
want the system (\ref{1}) to be nonlinear and coupled). Therefore, 
our definition  is by no means exhaustive: there exist  examples of 
multidimensional quasilinear systems which are integrable (in the 
inverse scattering sense) although do not possess n-component 
reductions. In all these examples the corresponding  matrices $A$ and 
$B$  commute: $[A, B]=0$.

{\bf Remark 3.} The anzatz somewhat similar to the key element of our 
construction was used for finding formal solutions of nonlinear 
evolution equations in the form of exponential series, see 
\cite{Hereman} and references therein, where solutions were sought as 
functions of real exponentials solving the linearized equation. The 
possibility to distinguish between integrable and nonintegrable 
equations in the framework of this approach was pointed out in 
\cite{Khus}.

In section 2 we discuss explicit examples which demonstrate that the 
above definition is indeed  very effective  for  detecting the 
integrability.

In section 3 we classify integrable systems of conservation laws in 
Godunov's form \cite{Godunov},
$$
v_t+(f_v)_x+(g_v)_y=0, ~~~ w_t+(f_w)_x+(g_w)_y=0;
$$
notice that systems of this type automatically possess one extra 
convex quadratic entropy
$$
\frac{1}{2}(v^2+w^2)_t+(vf_v+wf_w-f)_x+ (vg_v+wg_w-g)_y=0.
$$
The integrability conditions constitute a complicated overdetermined 
system (\ref{intf}) of fourth order PDEs for the potentials $f$ and 
$g$. The analysis of this system leads to two  possibilities.

\noindent {\bf Quadratic case.} There exists a linear combination of 
$f$ and $g$ which is quadratic in $v, w$. Without any loss of 
generality one can assume that $g$ is quadratic, say, $g=
(v^2-w^2)/2$ (one has a freedom of Euclidean isometries of the $(v, 
w)$-plane to bring $g$ to the canonical form). In this case our 
equations take the form
\begin{equation}
v_t+(f_v)_x+v_y=0, ~~~ w_t+(f_w)_x-w_y=0;
\label{q}
\end{equation}
the corresponding integrability conditions reduce to the system 
(\ref{f}) of  fourth order PDEs for the potential $f$ which can be 
solved explicitly (see Sect. 3). Notice that in the new independent 
variables
$\xi=-(t+y)/2, \  \eta=-(t-y)/2$ the system (\ref{q}) takes the form
$$
(f_v)_x=v_{\xi}, ~~~ (f_w)_x=w_{\eta},
$$
which is manifestly Hamiltonian in the new  variables $V=f_v, \ W=f_w$:
$$
V_x=(F_V)_{\xi}, ~~~ W_x=(F_W)_{\eta}.
$$
Here $F$ is the Legendre transform of $f$,  $F=vf_v+wf_w-f$. Thus, we 
obtain a complete description of the class of integrable 
two-component (2+1)-dimensional Hamiltonian systems of hydrodynamic 
type. An independent treatment of the Hamiltonian case is given in 
Sect. 4 where, in particular, the integrability conditions in terms 
of the Hamiltonian density $F$ are derived (formulae (\ref{4int})).

\noindent {\bf Harmonic case.} Here both $f$ and $g$ are 
harmonic functions. Further analysis shows that there exists a unique 
system of this type, with
$f=Re (z\ln z-z), \ g=Im (z\ln z-z), \ z=v+iw$. The corresponding equations are
$$
v_t+\frac{vv_x+ww_x}{v^2+w^2}+\frac{vw_y-wv_y}{v^2+w^2}=0, ~~~ 
w_t+\frac{wv_x-vw_x}{v^2+w^2}+\frac{vv_y+ww_y}{v^2+w^2}=0,
$$
or, in  conservative form,
$$
v_t+(\ln \sqrt {v^2+w^2})_x+(arctg \frac{w}{v})_y=0, ~~~ w_t-(arctg 
\frac{w}{v})_x+(\ln \sqrt {v^2+w^2})_y=0.
$$
Remarkably, this two-component system is a disguised form of the 
nonlinear wave equation. To see this we change to  polar coordinates 
$v=r\cos \theta, \ w=r \sin \theta$:
\begin{equation}
\theta_y=-\frac{r_x}{r}+r\sin \theta \ \theta _t-\cos \theta \ r_t, 
~~~ \theta_x=\frac{r_y}{r}+r\cos \theta \ \theta _t+\sin \theta \ r_t;
\label{B}
\end{equation}
these equations lead, upon cross-differentiation, to the nonlinear 
wave equation $(r^2)_{tt}=\triangle \ln (r^2)$, known also as the 
Boyer-Finley  equation \cite{Boyer}.
Solving equations (\ref{B}) for $r_x$ and $r_y$ and 
cross-differentiating, one obtains another second order equation 
$(r^2\theta_t)_t=\triangle \theta$ which can be viewed as a linear 
wave equation for $\theta$. It is not clear whether the wave equation 
with a general nonlinearity can be written as a two-component
first order system; however, as demonstrated in \cite{Keyfitz}, it 
always possesses a simple three-component representation.
Hydrodynamic reductions of the Boyer-Finley equation were recently 
studied in \cite{Fer}, see also Sect. 3.

systems  possessing infinitely many hydrodynamic reductions is 
provided in section 4. The corresponding Hamiltonian densities are 
solution space is 10-dimensional.

\section{Examples}

In this section we list some of the known examples of 
$(2+1)$-dimensional  systems of hydrodynamic type which are 
integrable in the above sense.
All these examples turn out to be conservative and, moreover,
possess exactly one `extra' conservation law which is the necessary 
ingredient of the theory of weak solutions. This makes the systems 
below a possible venue for  developing and testing mathematical 
theory (existence, uniqueness, weak solutions, etc.) of 
multidimensional conservation laws which,
currently, remain terra incognita \cite{Dafermos}.

{\bf Example 1.} The dispersionless KP equation,
$$
(u_t-uu_x)_x=u_{yy},
$$
plays an important role  in  nonlinear acoustics and differential 
geometry. Introducing the potential $u=\varphi_x$ we obtain the 
equation
$
\varphi_{xt}-\varphi_x\varphi_{xx}=\varphi_{yy}
$
which takes the quasilinear form
\begin{equation}
v_y=w_x, ~~~~ w_y=v_t-vv_x
\label{dKP}
\end{equation}
in the variables $v=\varphi_x, \ w=\varphi_y$.
Looking for reductions $v=v(R^1, ..., R^n), \ w=w(R^1, ..., R^n)$,
where the Riemann invariants $R^i$ satisfy  (\ref{R}), one readily obtains
\begin{equation}
\partial_i w=\mu^i \partial_i v, ~~~ \lambda^i=v+(\mu^i)^2.
\label{10}
\end{equation}
The compatibility condition 
$\partial_i\partial_jw=\partial_j\partial_iw$ implies
\begin{equation}
\partial_i\partial_jv=\frac{\partial_j\mu^i}{\mu^j-\mu^i}\partial_iv+\frac{\partial_i\mu^j}{\mu^i-\mu^j}\partial_jv,
\label{11}
\end{equation}
while the commutativity condition (\ref{comm}) results in
\begin{equation}
\partial_j\mu^i=\frac{\partial_j v}{\mu^j-\mu^i}.
\label{12}
\end{equation}
The substitution of (\ref{12}) into (\ref{11}) implies the 
Gibbons-Tsarev system for $v(R)$ and $\mu^i(R)$,
\begin{equation}
\partial_j\mu^i=\frac{\partial_j v}{\mu^j-\mu^i}, ~~~
\partial_i\partial_jv=2\frac{\partial_iv\partial_jv}{(\mu^j-\mu^i)^2},
\label{13}
\end{equation}
$i\ne j$, which was first derived in \cite{GibTsa96, GibTsa99} in the 
theory of hydrodynamic reductions of
Benney's moment equations, see also \cite{GuMaAl, Ma, Lei} for 
further discussion.
For any solution $\mu^i,  v$ of the system (\ref{13}) one can 
reconstruct $\lambda^i$ and $w$ by virtue of (\ref{10}).
In the two-component case  this system takes the form
\begin{equation}
\partial_2\mu^1=\frac{\partial_2 v}{\mu^2-\mu^1}, ~~~ 
\partial_1\mu^2=\frac{\partial_1 v}{\mu^1-\mu^2}, ~~~
\partial_1\partial_2v=2\frac{\partial_1v\partial_2v}{(\mu^2-\mu^1)^2}.
\label{14}
\end{equation}
The general solution of this system  is parametrized by four 
arbitrary functions of a single argument, indeed, one can arbitrarily 
prescribe the Goursat data $v(R^1), \ \mu^1(R^1)$ and   $v(R^2), \ 
\mu^2(R^2)$ on the characteristics $R^2=0$ and  $R^1=0$, 
respectively. Moreover, the system (\ref{14}) is invariant under the 
reparametrizations
$R^1\to f^1(R^1), \ R^2\to f^2(R^2)$ where $f^1, f^2$ are arbitrary 
functions of their arguments.
Since reparametrizations of this type do not effect the corresponding 
solutions, one concludes that two-component reductions are 
parametrized by two arbitrary  functions of a single argument.  A 
remarkable feature of the system (\ref{14}) is its multidimensional 
compatibility, that is, the compatibility of the system (\ref{13}) 
which is obtained by `gluing together' several identical copies of 
the system  (\ref{14}) for each pair of Riemann invariants $R^i, 
R^j$. Indeed, calculating $\partial_k(\partial_j\mu^i)$ by virtue of 
(\ref{13}) one obtains
$$
\partial_k(\partial_j\mu^i)=
\frac{\partial_jv\partial_kv(\mu^j+\mu^k-2\mu^i)}{(\mu^j-\mu^k)^2(\mu^j-\mu^i)(\mu^k-\mu^i)},
$$
which is manifestly symmetric in $j, k$. Therefore, the first 
compatibility condition
$\partial_k(\partial_j\mu^i)=\partial_j(\partial_k\mu^i)$ is 
satisfied. Similarly, the computation of
$\partial_k(\partial_i\partial_jv)$ results in
$$
\partial_k(\partial_i\partial_jv)=4\frac{\partial_iv \partial_jv \partial_kv
((\mu^i)^2+(\mu^j)^2+(\mu^k)^2-\mu^i\mu^j-\mu^i\mu^k-\mu^j\mu^k)}
{(\mu^i-\mu^j)^2(\mu^i-\mu^k)^2(\mu^j-\mu^k)^2},
$$
which is totally symmetric in $i, j, k$. Therefore, the second 
compatibility condition
$\partial_k(\partial_i\partial_jv)=\partial_j(\partial_i\partial_kv)$ 
is also satisfied. The general solution of the system (\ref{13}) 
depends, modulo trivial symmetries  $R^i\to f^i(R^i)$, on $n$ 
arbitrary functions of a single argument.

We just mention that  the system (\ref{dKP}) possesses exactly three 
conservation laws of hydrodynamic type:
$$
\begin{array}{c}
v_y=w_x, \\
\ \\
w_y=v_t-(v^2/2)_x, \\
\ \\
(vw)_y=(v^2/2)_t+(w^2/2-v^3/3)_x.
\end{array}
$$

{\bf Example 2.} The Boyer-Finley equation,
$$
u_{xy}=(e^u)_{tt},
$$
(notice that the signature here is different from the one discussed 
in the introduction, which makes the analysis much easier)  is 
descriptive of self-dual Einstein spaces with a Killing vector 
\cite{Boyer}. Introducing the potential $u=\varphi_t$, one obtains 
the equation
$
\varphi_{xy}=(e^{\varphi_t})_t
$
which takes the  form
\begin{equation}
v_t=w_x/w, ~~~~ w_t=v_y
\label{BF}
\end{equation}
in the new variables $v=\varphi_x, \ w=e^{\varphi_t}$.
Looking for reductions in the form $v=v(R^1, ..., R^n)$, $ w=w(R^1, ..., R^n)$,
where the Riemann invariants $R^i$ satisfy  (\ref{R}), one readily obtains
\begin{equation}
\partial_i w=w\lambda^i  \partial_i v, ~~~ \mu^i=w(\lambda^i)^2.
\label{20}
\end{equation}
The compatibility condition 
$\partial_i\partial_jw=\partial_j\partial_iw$ implies
\begin{equation}
\partial_i\partial_jv=\frac{\partial_j\lambda^i}{\lambda^j-\lambda^i}\partial_iv+\frac{\partial_i\lambda^j}{\lambda^i-\lambda^j}\partial_jv,
\label{21}
\end{equation}
while the commutativity condition (\ref{comm}) results in
\begin{equation}
\partial_j\lambda^i=\frac{(\lambda^i)^2\lambda^j 
}{\lambda^j-\lambda^i}\partial_j v.
\label{22}
\end{equation}
The substitution of (\ref{22}) into (\ref{21}) implies the   system 
for $v(R)$ and $\lambda^i(R)$,
\begin{equation}
\partial_j\lambda^i=\frac{(\lambda^i)^2\lambda^j 
}{\lambda^j-\lambda^i}\partial_j v, ~~~
\partial_i\partial_jv=\frac{\lambda^i \lambda^j(\lambda^i+\lambda^j) 
}{(\lambda^j-\lambda^i)^2}\partial_iv\partial_jv,
\label{23}
\end{equation}
which, in a somewhat different form,  was first discussed in \cite{Fer}.
For any solution $\lambda^i,  v$ of the system (\ref{23}) one can 
reconstruct $\mu^i, w$ by  virtue of  (\ref{20}). The system 
(\ref{23}) is compatible, with the general solution  depending on $n$ 
arbitrary functions of a single argument (modulo  trivial symmetries 
$R^i\to f^i(R^i)$).

The system (\ref{BF}) possesses three conservation laws of hydrodynamic type:
$$
\begin{array}{c}
w_t=v_y, \\
\ \\
v_t=(\ln w)_x, \\
\ \\
(vw)_t=w_x+(v^2/2)_y.
\end{array}
$$

{\bf Example 3.} An interesting integrable modification of the 
Boyer-Finley equation is the PDE
$$
u_{xy}=(\partial_t^2-c\partial_x^2)(e^u),
$$
$c=const$ (see \cite{Pavlov, Pavlov1} for further examples of this 
type). Introducing the potential $e^u=\varphi_x$, one obtains the 
equation
$
\varphi_{tt}-c\varphi_{xx}=(\ln {\varphi_x})_y
$
which takes the form
\begin{equation}
v_t=cw_x+ w_y/w, ~~~~ w_t=v_x
\label{deform}
\end{equation}
in the new variables $v=\varphi_t, \ w=\varphi_x$.
Looking for reductions in the form $v=v(R^1, ..., R^n)$, $ w=w(R^1, ..., R^n)$,
where the Riemann invariants $R^i$ satisfy  (\ref{R}), one  obtains
\begin{equation}
\partial_i v=\lambda^i  \partial_i w, ~~~ \mu^i=w((\lambda^i)^2-c).
\label{30}
\end{equation}
The compatibility condition 
$\partial_i\partial_jv=\partial_j\partial_iv$ implies
\begin{equation}
\partial_i\partial_jw=\frac{\partial_j\lambda^i}{\lambda^j-\lambda^i}\partial_iw+\frac{\partial_i\lambda^j}{\lambda^i-\lambda^j}\partial_jw,
\label{31}
\end{equation}
while the commutativity condition (\ref{comm}) results in
\begin{equation}
\partial_j\lambda^i=\frac{(\lambda^i)^2-c}{\lambda^j-\lambda^i}\frac{ 
\partial_j w}{w}.
\label{32}
\end{equation}
The substitution of (\ref{32}) into (\ref{31}) implies the  following 
system for $w(R)$ and $\lambda^i(R)$,
\begin{equation}
\partial_j\lambda^i=\frac{(\lambda^i)^2-c}{\lambda^j-\lambda^i} 
\frac{\partial_j w}{w}, ~~~
\partial_i\partial_jw=\frac{(\lambda^i)^2+(\lambda^j)^2-2c}{w(\lambda^j-\lambda^i)^2}\partial_iw\partial_jw.
\label{33}
\end{equation}
For any solution $\lambda^i,  w$ of the system (\ref{33}) one can 
reconstruct $\mu^i, v$ by  virtue of  (\ref{30}). The system 
(\ref{33}) is compatible, with the general solution  depending on $n$ 
arbitrary functions of a single argument (modulo   symmetries $R^i\to 
f^i(R^i)$).

The system (\ref{deform}) possesses three conservation laws of 
hydrodynamic type:
$$
\begin{array}{c}
v_t=cw_x+(\ln w)_y, \\
\ \\
w_t=v_x, \\
\ \\
(vw)_t=(v^2/2+w^2/(2c))_x+w_y.
\end{array}
$$
Notice that  when $c>0$ the $x$-flux of the third conservation law is 
a convex function of the previous two $x$-fluxes (convex entropy). 
This example can be of particular interest for the general theory of 
multidimensional strictly hyperbolic conservation laws.

{\bf Example 4.} We are going to demonstrate that the method of 
hydrodynamic reductions is in fact the effective classification 
criterion. As an illustration of our approach we will classify 
integrable nonlinear wave equations of the form
$$
u_{xy}=(f(u))_{tt};
$$
it will follow that $f(u)=e^u$ is the only nontrivial possibility. 
Introducing the potential $u=\varphi_t$, one obtains the equation
$
\varphi_{xy}=(f({\varphi_t}))_t
$
which takes the  form
\begin{equation}
v_t=w_x, ~~~~ f'(w) w_t=v_y
\label{wave}
\end{equation}
in the new variables $v=\varphi_x, \ w=\varphi_t$.
Looking for reductions in the form $v=v(R^1, ..., R^n)$, $ w=w(R^1, ..., R^n)$
where the Riemann invariants $R^i$ satisfy  (\ref{R}) one readily obtains
\begin{equation}
\partial_i v= \partial_i w/\lambda^i, ~~~ \mu^i=f'(w)(\lambda^i)^2.
\label{40}
\end{equation}
The compatibility condition 
$\partial_i\partial_jv=\partial_j\partial_iv$ implies
\begin{equation}
\partial_i\partial_jw=\frac{\lambda^j\partial_j\lambda^i}{(\lambda^j-\lambda^i)\lambda^i}\partial_iw+\frac{\lambda^i\partial_i\lambda^j}{(\lambda^i-\lambda^j)\lambda^j}\partial_jw,
\label{41}
\end{equation}
while the commutativity condition (\ref{comm}) results in
\begin{equation}
\partial_j\lambda^i=\frac{f''}{f'}\frac{(\lambda^i)^2}{\lambda^j-\lambda^i}\partial_j 
w.
\label{42}
\end{equation}
The substitution of (\ref{22}) into (\ref{21}) implies the   system 
for $w(R)$ and $\lambda^i(R)$,
\begin{equation}
\partial_j\lambda^i=\frac{f''}{f'}\frac{(\lambda^i)^2}{\lambda^j-\lambda^i}\partial_j 
w, ~~~
\partial_i\partial_jw=2\frac{f''}{f'}\frac{ \lambda^i 
\lambda^j}{(\lambda^j-\lambda^i)^2}\partial_iw\partial_jw.
\label{43}
\end{equation}
A direct computation of $\partial_k(\partial_j\lambda^i)$ implies
$$
\begin{array}{c}
\partial_k(\partial_j\lambda^i)=\left(\frac{f''}{f'}\right)'\frac{(\lambda^i)^2}{\lambda^j-\lambda^i}\partial_jw\partial_kw+ 
\\
\ \\
\left(\frac{f''}{f'}\right)^{2}\frac{(\lambda^i)^2
(\lambda^i((\lambda^j)^2+(\lambda^k)^2)+\lambda^j\lambda^k(\lambda^j+\lambda^k)-4\lambda^i\lambda^j\lambda^k)}
{(\lambda^j-\lambda^k)^2(\lambda^j-\lambda^i)(\lambda^k-\lambda^i)}\partial_jw\partial_kw.
\end{array}
$$
The compatibility condition 
$\partial_k(\partial_j\lambda^i)=\partial_j(\partial_k\lambda^i)$
is equivalent to the requirement that the above expression is 
symmetric in $j, k$. Since the second term is manifestly symmetric, 
one has to require
$$
\left(\frac{f''}{f'}\right)'=0
$$
to ensure the compatibility. This implies  $f=ae^{bu}+c$, which is 
essentially the case of Example 2.

{\bf Example 5.} There exist remarkable examples with a fairly simple 
structure of hydrodynamic reductions. Let us consider the system 
\cite{Pavlov}
\begin{equation}
v_t=w_x, ~~~~ w_t=wv_x-vw_x+v_y.
\label{max}
\end{equation}
Looking for reductions in the form $v=v(R^1, ..., R^n), \ w=w(R^1, ..., R^n)$
where the Riemann invariants $R^i$ satisfy  (\ref{R}) one obtains
$$
\partial_i w=\lambda^i \partial_iv, ~~~ \mu^i=(\lambda^i)^2+v\lambda^i-w,
$$
so that
$$
\partial_i\partial_jv=\frac{\partial_j\lambda^i}{\lambda^j-\lambda^i}\partial_iv+\frac{\partial_i\lambda^j}{\lambda^i-\lambda^j}\partial_jv.
$$
The commutativity condition (\ref{comm}) implies
$$
\partial_j\lambda^i=-\partial_jv, ~~~ \partial_i\partial_jv=0.
$$
Hence,
$$
v=\sum_k f^k(R^k), ~~~ \lambda^i=\varphi^i(R^i)-\sum_k f^k(R^k), \\
$$
where $f^i(R^i)$ and $\varphi^i(R^i)$ are arbitrary functions of a 
single argument. Further properties of these reductions were 
investigated  in \cite{Pavlov}. We just mention that  the system 
(\ref{max}) possesses  three conservation laws of hydrodynamic type:
$$
\begin{array}{c}
v_t=w_x, \\
\ \\
(w+v^2)_t=(vw)_x+v_y, \\
\ \\
(2vw+v^3)_t=(v^2w+w^2)_x+(v^2)_y.
\end{array}
$$

{\bf Example 6.} The method of hydrodynamic reductions carries over 
to multicomponent situation in a straightforward way. Here we give 
details of calculations for the 3-component system first proposed in 
\cite{Zakharov}, see also \cite{GuMaAl}:
\begin{equation}
a_t+(av)_x=0, ~~~ v_t+vv_x+w_x=0, ~~~ w_y+a_x=0.
\label{Z}
\end{equation}
Looking for reductions in the form $a=a(R^1, ..., R^n)$, $ b=b(R^1, 
..., R^n)$,  $ w=w(R^1, ..., R^n)$
where the Riemann invariants $R^i$ satisfy  (\ref{R}) one  obtains 
the relations
\begin{equation}
\partial_i w=-(\lambda^i+v)  \partial_i v, ~~~ \partial_i a=\mu^i 
(\lambda^i+v)  \partial_i v, ~~~ \mu^i=-\frac{a}{(\lambda^i+v)^2}.
\label{60}
\end{equation}
The compatibility condition 
$\partial_i\partial_jw=\partial_j\partial_iw$ implies
\begin{equation}
\partial_i\partial_jv=\frac{\partial_j\lambda^i}{\lambda^j-\lambda^i}\partial_iv+\frac{\partial_i\lambda^j}{\lambda^i-\lambda^j}\partial_jv,
\label{61}
\end{equation}
while the commutativity condition (\ref{comm}) results in
\begin{equation}
\partial_j\lambda^i=\frac{\lambda^j+v}{\lambda^i-\lambda^j}\partial_j v.
\label{62}
\end{equation}
The substitution of (\ref{62}) into (\ref{61}) implies the   system 
for $v(R)$ and $\lambda^i(R)$,
\begin{equation}
\partial_j\lambda^i=\frac{\lambda^j+v}{\lambda^i-\lambda^j}\partial_j v, ~~~
\partial_i\partial_jv=-\frac{\lambda^i+\lambda^j+2v}{(\lambda^j-\lambda^i)^2}\partial_iv\partial_jv 
.
\label{63}
\end{equation}
One can verify that the remaining compatibility conditions 
$\partial_i\partial_ja=\partial_j\partial_ia$ are satisfied 
identically.
For any solution $\lambda^i,  v$ of the system (\ref{63}) one can 
reconstruct $w, a, \mu^i$ by  virtue of  (\ref{60}). The system 
(\ref{63}) is compatible, with the general solution  depending on $n$ 
arbitrary functions of a single argument (modulo  symmetries $R^i\to 
f^i(R^i)$).

Notice that the system (\ref{Z}) possesses four conservation laws of 
hydrodynamic type:
$$
\begin{array}{c}
a_t+(av)_x=0, \\
\ \\
v_t+(v^2/2+w)_x=0, \\
\ \\
w_y+a_x=0,\\
\ \\
(aw+av^2)_x+(w^2/2)_y+(av)_t=0.
\end{array}
$$
Equations (\ref{Z}) can be generalized as follows:
$$
a_t+(av)_x=0, ~~~ v_t+vv_x+w_x=0, ~~~ w_y+p(a)_x=0.
$$
(so that one obtains isentropic gas dynamics in the limit $x=y$). It 
can be shown that this system passes the  integrability test  if and 
only if
$p''=0$, which leads to  (\ref{Z}).

\section{Classification of  integrable  systems of conservation laws 
with a convex quadratic entropy}

In this section we discuss  systems of conservation laws  in 
Godunov's form \cite{Godunov},
\begin{equation}
v_t+(f_v)_x+(g_v)_y=0, ~~~ w_t+(f_w)_x+(g_w)_y=0;
\label{31a}
\end{equation}
here $f(v, w)$ and $g(v, w)$ are given potentials. Systems of this 
type automatically possess one extra convex quadratic entropy
$$
\frac{1}{2}(v^2+w^2)_t+(vf_v+wf_w-f)_x+ (vg_v+wg_w-g)_y=0.
$$
Equations (\ref{31a}) can be written in the matrix form (\ref{1}) with
$$
{\bf u}=\left(\begin{array}{c}v\\
w
\end{array}\right), ~~~ A=\left(\begin{array}{cc}f_{vv}&f_{vw}\\
f_{vw}&f_{ww}
\end{array}\right), ~~~  B=\left(\begin{array}{cc}g_{vv}&g_{vw}\\
g_{vw}&g_{ww}
\end{array}\right);
$$
in what follows we assume that the commutator
$$
[A, B]= \left(\begin{array}{cc} 0&s\\
-s&0
\end{array}\right)
$$
is nonzero, that is, 
$s=g_{vw}(f_{vv}-f_{ww})-f_{vw}(g_{vv}-g_{ww})\ne 0$; otherwise, the 
system  possesses no nontrivial $n$-component hydrodynamic reductions.
The integrability conditions   lead to an overdetermined system of 
fourth order PDEs for $f$ and $ g$. A careful analysis of this system 
shows that there exist two essentially different possibilities:

\noindent --   $g$  is  quadratic in $v, w$ (quadratic case) or

\noindent -- both  $f$ and $g$ are harmonic functions  (harmonic case).

\noindent  Remarkably, in both cases the equations for $f$  and $g$ can 
be solved in a closed form. We hope that these examples  would 
provide a good venue for developing and testing the general theory of 
multidimensional conservation laws (breakdown of solutions, weak 
solutions, etc).


The integrability conditions  can be derived in the standard way. 
Looking for reductions of the system (\ref{31}) in the form $v=v(R^1, 
..., R^n)$, $w=w(R^1, ..., R^n)$ where the Riemann invariants satisfy 
equations (\ref{R}), and substituting into (\ref{31}), one  arrives at
$$
(\lambda^i+f_{vv}+\mu^ig_{vv})\partial_iv+(f_{vw}+\mu^ig_{vw})\partial_iw=0, 
~~~
(f_{vw}+\mu^ig_{vw})\partial_iv+(\lambda^i+f_{ww}+ \mu^ig_{ww})\partial_iw=0,
$$
(no summation!) so that $\lambda^i$ and $\mu^i$ satisfy the dispersion relation
$$
(\lambda^i+f_{vv}+\mu^ig_{vv})(\lambda^i+f_{ww}+ 
\mu^ig_{ww})=(f_{vw}+\mu^ig_{vw})^2.
$$
  Setting $\partial_iv=\varphi^i\partial_iw$ one obtains the following 
expressions for $\lambda^i$ and $\mu^i$ in terms of $\varphi^i$,
$$
\begin{array}{c}
\displaystyle{\lambda^i=\frac{(f_{vv}g_{vw}-f_{vw}g_{vv})(\varphi^i)^2+(f_{vv}g_{ww}-f_{ww}g_{vv})\varphi^i+(f_{vw}g_{ww}-f_{ww}g_{vw})}{g_{vw}(1-(\varphi^i)^2)+(g_{vv}-g_{ww})\varphi^i}}, 
\\
\ \\
\displaystyle{\mu^i=-\frac{f_{vw}(1-(\varphi^i)^2)+(f_{vv}-f_{ww})\varphi^i}{g_{vw}(1-(\varphi^i)^2)+(g_{vv}-g_{ww})\varphi^i}},
\end{array}
$$
which define a rational parametrization of the dispersion relation. 
The compatibility conditions of the equations 
$\partial_iv=\varphi^i\partial_iw$ imply
\begin{equation}
\partial_i\partial_jw=\frac{\partial_j\varphi^i}{\varphi^j-\varphi^i}\partial_iw+\frac{\partial_i\varphi^j}{\varphi^i-\varphi^j}\partial_jw,
\label{3w}
\end{equation}
while the commutativity conditions (\ref{comm}) lead to the 
expressions for $\partial_j \varphi^i$ in the form $\partial_j 
\varphi^i=(...)\partial_jw$. Here dots denote a rational expression 
in $\varphi^i, \varphi^j$
whose coefficients are  functions of the second and third derivatives 
of $f$ and $g$. We do not write them  down explicitly due to their 
complexity. To manipulate with these expressions we used symbolic 
computations.
One can  see that  the  compatibility condition 
$\partial_k\partial_j\varphi^i-\partial_j\partial_k\varphi^i=0$ is of 
the form
$P\partial_jw\partial_kw=0$, where $P$ is a complicated rational 
expression in $\varphi^i, \varphi^j, \varphi^k$ whose coefficients 
depend on partial derivatives of $f$ and $g$ up to fourth order.
Requiring that  $P$ vanishes identically we obtain  the 
overdetermined system of fourth order PDEs  for $f$ and $g$. This 
system yields the following expressions for the fourth derivatives of 
$f$:
\begin{eqnarray}
sf_{vvvv}&=&f_{vvv}[2(g_{ww}-g_{vv})f_{vvw}+3(f_{vv}-f_{ww})g_{vvw}+ 
\nonumber \\
&&2g_{vw}(f_{vvv}+2f_{vww})-2f_{vw}g_{vww}] +\nonumber  \\
&& g_{vvv}[(f_{ww}-f_{vv})f_{vvw}-2f_{vw}(f_{vww}+f_{vvv})] +\nonumber \\
&&6f_{vvw}(f_{vw}g_{vvw}-g_{vw}f_{vvw}),  \nonumber \\
-sf_{vvvw}&=&f_{vvw}(f_{vw}g_{vvv}+3g_{vw}f_{vww}-3f_{vw}g_{vww})+ 
f_{vvv}[2(f_{ww}-f_{vv})g_{vww}-\nonumber \\
&&2(g_{ww}- 
g_{vv})f_{vww}+f_{vw}(g_{www}+g_{vvw})-g_{vw}(f_{www}+2f_{vvw})], 
\nonumber \\
sf_{vvww}&=&(g_{ww}-g_{vv})(f_{vww}f_{vvw}+f_{vvv}f_{www})+\nonumber \\
&&(f_{vv}-f_{ww})(f_{vww}g_{vvw}+g_{vvv}f_{www})+ \label{intf} \\
&& 
2g_{vw}(f_{vvw}^2-f_{vww}^2)+2f_{vw}(f_{vww}g_{vww}-f_{vvw}g_{vvw}),\nonumber 
\\
sf_{vwww}&=&f_{vww}(f_{vw}g_{www}+3g_{vw}f_{vvw}-3f_{vw}g_{vvw})+
f_{www}[2(f_{vv}-f_{ww})g_{vvw}- \nonumber \\
&&2(g_{vv}-g_{ww})f_{vvw}+f_{vw}(g_{vvv}+g_{vww})-
g_{vw}(f_{vvv}+2f_{vww})], \nonumber \\
-sf_{wwww}&=&f_{www}[2(g_{vv}-g_{ww})f_{vww}+3(f_{ww}-f_{vv})g_{vww}+ 
\nonumber \\
&&2g_{vw}(f_{www}+2f_{vvw})-2f_{vw}g_{vvw}]+ \nonumber \\
&&g_{www}[(f_{vv}-f_{ww})f_{vww}-2f_{vw}(f_{vvw}+f_{www})] + \nonumber \\
&&6f_{vww}(f_{vw}g_{vww}-g_{vw}f_{vww}); \nonumber
\end{eqnarray}
here $s=g_{vw}(f_{vv}-f_{ww})-f_{vw}(g_{vv}-g_{ww})\ne 0$. Notice 
that  equations $(\ref{intf})_4$ and $(\ref{intf})_5$ can be obtained 
from $(\ref{intf})_2$ and $(\ref{intf})_1$ by interchanging $v$ and 
$w$. Analogous expressions for the fourth derivatives of $g$ can be 
obtained from (\ref{intf}) by interchanging $f$ and $g$.
Moreover, one has five quadratic relations among the  third 
derivatives of $f$ and $g$:
\begin{equation}
\begin{array}{c}
f_{vvv}g_{www}-f_{www}g_{vvv}+f_{vww}g_{www}-f_{www}g_{vww}=0, \\
f_{vvv}g_{www}-f_{www}g_{vvv}+f_{vvv}g_{vvw}-f_{vvw}g_{vvv}=0, \\
f_{vvv}g_{www}-f_{www}g_{vvv}+f_{vvw}g_{vww}-f_{vww}g_{vvw}=0, \\
f_{vvv}g_{vww}-f_{vww}g_{vvv}=0, \\
  f_{vvw}g_{www}-f_{www}g_{vvw}=0.
\end{array}
\label{quadfg}
\end{equation}
This system of PDEs for $f$ and $g$ is not in involution;
a careful analysis of the quadratic relations (\ref{quadfg}) leads to 
two essentially different possibilities:

\noindent {\bf Quadratic case.} The third derivatives of $g$ are 
proportional to the corresponding third derivatives of $f$,
$$
g_{vvv}=\mu f_{vvv}, ~~ g_{vvw}=\mu f_{vvw}, ~~ g_{vww}=\mu f_{vww}, 
~~ g_{www}=\mu f_{www}.
$$
Substituting these relations into the remaining integrability 
conditions  one can show that $\mu$ must be constant. Therefore, 
$g-\mu f$ is at most quadratic in $v, w$.
Without any loss of generality one can assume that, say, $g$ is quadratic.

\noindent {\bf Harmonic case.}  Here
$$
f_{vvv}+f_{vww}=0, ~~  f_{www}+f_{vvw}=0, ~~ g_{vvv}+g_{vww}=0, ~~ 
g_{www}+g_{vvw}=0,
$$
which imply that $\triangle f$ and $\triangle g$ are constants. 
Without any loss of generality one can assume that both $f$ and $g$ 
are harmonic, $\triangle f=\triangle g=0$.

{\bf Remark 4.} There exists an obvious group of equivalence 
transformations which preserve the integrability and leave equations 
(\ref{31a}) form-invariant. These are, first of all,
orthogonal  transformations of the $(v, w)$-plane, generated by 
translations and  rotations.  Secondly, these are linear changes of 
the independent variables in (\ref{31a}),
$$
x\to a_{11}x+a_{12}y+a_{13}t, ~~~ y\to a_{21}x+a_{22}y+a_{23}t,
$$
which induce the transformations
$$
f\to a_{11}f+a_{12}g+a_{13}\frac{v^2+w^2}{2}, ~~~ g\to 
a_{21}f+a_{22}g+a_{23}\frac{v^2+w^2}{2}.
$$
The classification below is carried out up to this natural equivalence.

\subsection{Quadratic case}
Setting $g(v, w)=av^2+2bvw+cw^2$ and introducing 
$C=b(f_{ww}-f_{vv})+(a-c)f_{vw}$  one can rewrite equations 
(\ref{intf})  as follows:
\begin{equation}
\begin{array}{c}
Cf_{vvvv}=2b(3f^2_{vvw}-2f_{vvv}f_{vww}-f_{vvv}^2)+2(a-c)f_{vvv}f_{vvw}, \\
\ \\
Cf_{vvvw}=b(3f_{vvw}f_{vww}-2f_{vvv}f_{vvw}-f_{vvv}f_{www})+2(a-c)f_{vvv}f_{vww}, 
\\
\ \\
Cf_{vvww}=2b(f_{vww}^2-f_{vvw}^2)+(a-c)(f_{vvw}f_{vww}+f_{vvv}f_{www}), \\
\ \\
Cf_{vwww}=-b(3f_{vww}f_{vvw}-2f_{www}f_{vww}-f_{vvv}f_{www})+2(a-c)f_{vvw}f_{www}, 
\\
\ \\
Cf_{wwww}=-2b(3f^2_{vww}-2f_{www}f_{vvw}-f_{www}^2)+2(a-c)f_{vww}f_{www},
\end{array}
\label{3int}
\end{equation}
(the corresponding equations for $g$ are satisfied identically). For 
any   $a, b, c$ this system is in involution and its solution space 
is 10-dimensional. Indeed, the values of $f$, its first, second and 
third derivatives can be choosen arbitrarily, while the fourth and 
higher derivatives are determined by virtue of (\ref{3int}). 
Diagonalizing the quadratic form $g$ by  a linear orthogonal change 
of variables $v, w$, one can set $b=0$. In this case equations
(\ref{3int}) simplify to
\begin{equation}
\begin{array}{c}
f_{vw}f_{vvvv}=2f_{vvv}f_{vvw}, \\
\ \\
f_{vw}f_{vvvw}=2f_{vvv}f_{vww}, \\
\ \\
f_{vw}f_{vvww}=f_{vvw}f_{vww}+f_{vvv}f_{www}, \\
\ \\
f_{vw}f_{vwww}=2f_{vvw}f_{www}, \\
\ \\
f_{vw}f_{wwww}=2f_{vww}f_{www}.
\end{array}
\label{f}
\end{equation}
The first two  equations  imply that $f_{vvv}/f_{vw}^2={\rm const}$. 
Similarly, the last two equations imply $f_{www}/f_{vw}^2={\rm 
const}$.
Setting $f_{vw}=e$ one can parametrise the third derivatives of $f$ as follows:
\begin{equation}
f_{vvv}=\frac{1}{2}me^2, ~~~ f_{vvw}=e_v, ~~~ f_{vww}=e_w, ~~~ 
f_{www}=\frac{1}{2}ne^2;
\label{3rd}
\end{equation}
here $m, n $ are arbitrary constants. The compatibility conditions 
of these equations plus the equation $(\ref{f})_3$  result 
in the following  overdetermined system for $e$:
\begin{equation}
(\ln e)_{vw}=\frac{mn}{4}e^2, ~~~ e_{vv}=mee_{w}, ~~~ e_{ww}=nee_v.
\label{e1}
\end{equation}
It is worth mentioning that the  system (\ref{e1}) arises in a 
completely different context  in  projective differential geometry 
constituting the projective Gauss-Codazzi equations of the Roman 
surface of Steiner  which is known as the only quartic in $P^3$ 
containing a two-parameter family of conics, see e.g. 
\cite{proj}.
At the moment we have no explanation of this remarkable coincidence.

\noindent Solving the first (Liouville) equation for $e$ in the form
$$
e^2=\frac{4}{mn}\frac{p'(v)q'(w)}{(p(v)+q(w))^2}
$$
and setting
\begin{equation}
(p')^{3/2}=\sqrt {m}\ P(p), ~~~ (q')^{3/2}=\sqrt{n}\ Q(q),
\label{pq}
\end{equation}
(here $P(p)$ and $Q(q)$ are functions to be determined),  one obtains 
from the last two equations (\ref{e1}) the following 
functional-differential equations for $P$ and $Q$:
$$
P''(p+q)^2-4P'(p+q)+6P=2Q'(p+q)-6Q,  ~~~ 
Q''(p+q)^2-4Q'(p+q)+6Q=2P'(p+q)-6P;
$$
these equations  imply that 
both $P$ and $Q$ are cubic polynomials in $p$ and $q$,
$$
P=ap^3+bp^2+cp+d, ~~~ Q=aq^3-bq^2+cq-d,
$$
where $a, b, c, d$ are arbitrary constants.  In the general case, 
using translations and scalings, 
equations (\ref{pq}) can be brought to the form $(y')^3 = (y^3 - 3 \lambda y^2 + 3 y)^2$
and solved in terms of elliptic functions 
\cite{Abram}:
$$
y = \frac{2}{\lambda - 3 \wp' (z; 0, g_3)}, \quad 
g_3 = \frac{4-3 \lambda^2}{27}.
$$

In the simplest case $m=n=0$ equations (\ref{e1}) imply
$$
e=(\alpha v +\beta)( \gamma w +\delta),
$$
and the elementary integration of  (\ref{3rd}) results in
$$
f(v, w)=\frac{\alpha \gamma}{4} v^2w^2+\frac{\alpha \delta}{2} 
v^2w+\frac{\beta \gamma}{2} vw^2+\beta \delta vw;
$$
here $\alpha, \beta, \gamma, \delta$ are arbitrary constants. Using 
the equivalence transformations  one can reduce $f$ 
to either $f=v^2w^2$ (if both $\alpha$ and $\gamma$ are nonzero) or 
$f=vw^2$ (if $\alpha =0$). The corresponding equations (\ref{31a}) 
take the form
$$
v_t+2(vw^2)_x+v_y=0, ~~~ w_t+2(v^2w)_x-w_y=0
$$
and
$$
v_t+(w^2)_x+v_y=0, ~~~ w_t+2(vw)_x-w_y=0,
$$
respectively (in both cases $g=(v^2-w^2)/2$).

\noindent If $m=0, \ n\ne 0$, equations (\ref{e1}) imply
$$
e=(\alpha v +\beta)\varphi '(w)
$$
where $\varphi '''=\alpha n (\varphi ')^2, \ \alpha, \beta = 
const$.
Therefore, $\varphi (w) = - \frac{6}{\alpha n} \zeta (w; 0, 
g_3) + \gamma$, where  $\zeta'(z) = -\wp (z)$ 
and $g_3, \gamma$ are constants. 
The elementary integration of the equations (\ref{3rd}) gives
$$
f=\frac{1}{2\alpha }(\alpha v+\beta)^2\varphi (w).
$$
This reduces to the previous case if $n=0$.

\subsection{Harmonic case}
We have 
$f_{vv}=-f_{ww}, \ g_{vv}=-g_{ww}$, so that
  $$
f_{vvv}=-f_{vww},  ~~~ f_{www}=-f_{vvw},  ~~~ g_{vvv}=-g_{vww},  ~~~ 
g_{www}=-g_{vvw}.
$$
Differentiating these relations by virtue of (\ref{intf}) (and the 
analogous equations for $g$) one arrives at the  additional 
constraints
$$
\begin{array}{c}
f_{vv}p_1+f_{vw}p_2+g_{ww}p_3=0, ~~~ -f_{vw}p_1+f_{vv}p_2+g_{vw}p_3=0,\\
\ \\
-g_{vw}p_1+g_{ww}p_2+f_{vw}p_4=0, ~~~ g_{ww}p_1+g_{vw}p_2+f_{vv}p_4=0
\end{array}
$$
where
$$
p_1=f_{vww}g_{vww}+f_{vvw}g_{vvw}, ~~~ 
p_2=f_{vww}g_{vvw}-f_{vvw}g_{vww}, ~~~ p_3=f^2_{vvw}+f^2_{vww}, ~~~ 
p_4=g^2_{vvw}+g^2_{vww}.
$$
Solving the linear system for $p_i$ (notice that the corresponding 
$4\times 4$ matrix has rank three) one obtains
\begin{equation}
\begin{array}{c}
p_1=f_{vww}g_{vww}+f_{vvw}g_{vvw}=\mu^2(f_{vw}g_{vw}-f_{vv}g_{ww}),\\
  \ \\
p_2=f_{vww}g_{vvw}-f_{vvw}g_{vww}=-\mu^2(f_{vv}g_{vw}+f_{vw}g_{ww}), \\
\ \\
p_3=f^2_{vvw}+f^2_{vww}=\mu^2(f_{vw}^2+f_{vv}^2), \\
\ \\
  p_4=g^2_{vvw}+g^2_{vww}=\mu^2(g_{vw}^2+g_{ww}^2).
\end{array}
\label{vect}
\end{equation}
Introducing the two-component vectors
$$
e_1=\left(\begin{array}{c}f_{vww} \\ f_{vvw} \end{array}\right), ~~~ 
e_2=\left(\begin{array}{c}g_{vww} \\ g_{vvw}\end{array}\right), ~~~ 
s_1=\left(\begin{array}{c}f_{vw} \\ f_{ww}\end{array}\right), ~~~ 
s_2=\left(\begin{array}{c}g_{vw} \\ g_{ww}\end{array}\right),
$$
one can rewrite (\ref{vect}) in vector notation as follows
$$
(e_1, e_1)=\mu^2(s_1, s_1), ~~~ (e_2, e_2)=\mu^2(s_2, s_2), ~~~ (e_1, 
e_2)=\mu^2(s_1, s_2), ~~~ e_1\wedge  e_2=-\mu^2 s_1\wedge s_2
$$
where $( , )$ is the standard Euclidean scalar product. Therefore, 
the vectors $e_1, e_2$ and $s_1, s_2$ are related by a composition of 
a scaling, rotation and reflection, that is,
$$
e_1= \left(\begin{array}{cc}
X&Y\\
Y&-X
\end{array} \right) s_1, ~~~
e_2= \left(\begin{array}{cc}
X&Y\\
Y&-X
\end{array} \right) s_2
$$
or, explicitly,
\begin{equation}
\begin{array}{c}
f_{vww}=Xf_{vw}+Yf_{ww}, ~~~ f_{vvw}=Yf_{vw}-Xf_{ww}, \\
  g_{vww}=Xg_{vw}+Yg_{ww}, ~~~ g_{vvw}=Yg_{vw}-Xg_{ww}
\end{array}
\label{harm}
\end{equation}
(notice that equations for $f$ and $g$ coincide). The compatibility 
conditions of these equations imply the following equations for $X$ 
and $Y$,
$$
X_v=2XY, ~~~ X_w=X^2-Y^2, ~~~ Y_v=Y^2-X^2, ~~~ Y_w=2XY,
$$
whose general solution (up to translations in $v$ and $w$) is
$$
X=-\frac{w}{v^2+w^2}, ~~~ Y=-\frac{v}{v^2+w^2}.
$$
Substituting these expressions into (\ref{harm}) and integrating the 
corresponding linear system for the harmonic function $f$ (this 
integration simplifies if one changes to the complex variables
$z=v+iw, \ \bar z=v-iw$) one readily obtains that $f$ must be a 
linear combination of the real and imaginary parts of the
function $z\ln z-z$. The same result holds for $g$.
Therefore, without any loss of generality one can set $f=Re (z\ln 
z-z), \ g=Im (z\ln z-z)$.

\subsection{Reductions of the system $\bf (\ref{B})$}

Here we briefly discuss reductions of the system (\ref{B}),
$$
\theta_y=-\frac{r_x}{r}+r\sin \theta \ \theta _t-\cos \theta \ r_t, 
~~~ \theta_x=\frac{r_y}{r}+r\cos \theta \ \theta _t+\sin \theta \ r_t.
$$
Assuming  $\theta=\theta(R^1, ..., R^n)$, $r=r(R^1, ..., R^n)$ where 
the Riemann invariants satisfy the equations
$$
R^i_x=\lambda^i(R)\ R^i_t, ~~~~ R^i_y=\mu^i(R)\ R^i_t
$$
and substituting into (\ref{B}) one  arrives at
$$
(\mu^i-r\sin \theta)\partial_i\theta+(\lambda^i/r+\cos 
\theta)\partial_ir=0, ~~~
(\lambda^i-r\cos \theta)\partial_i\theta-(\mu^i/r+\sin \theta)\partial_ir=0,
$$
(no summation!) Hence, the characteristic speeds $\lambda^i$ and 
$\mu^i$ satisfy the dispersion relation which  takes a particularly 
simple form
$$
(\lambda^i)^2+(\mu^i)^2=r^2.
$$
Parametrising $\lambda^i$ and $\mu^i$ in the form $\lambda^i=r\cos 
\varphi^i, \ \mu^i=r\sin \varphi^i$ we obtain
\begin{equation}
\partial_i \theta=\frac{\cos \theta +\cos \varphi^i}{\sin \theta - 
\sin \varphi^i}\ \partial_iU,
\label{t}
\end{equation}
$U=\ln r$. The compatibility conditions of these equations together with the commutativity conditions (\ref{comm}) imply 
the  system
\begin{equation}
\partial_i\partial_jU=-\frac{\partial_iU\partial_jU}{ 
\sin^2\frac{\varphi^i-\varphi^j}{2}}, ~~~ \partial_j\varphi^i=\cot 
\frac{\varphi^i-\varphi^j}{2} \ \partial_jU
\label{tt}
\end{equation}
which is a trigonometric version of the Gibbons-Tsarev system 
(\ref{13}). As shown in \cite{Fer} this system is, in fact, 
equivalent to (\ref{13}), and its solutions can be constructed from 
the known solutions of the Gibbons-Tsarev system \cite{GibTsa96, 
GibTsa99}. Once the solution of (\ref{tt}) is known, the 
corresponding polar angle $\theta$ can be calculated from the 
equations (\ref{t}) which are compatible by construction.

\section{Classification of integrable  Hamiltonian systems  of 
hydrodynamic type in $2+1$ dimensions}

In this section we classify  Hamiltonian systems
\begin{equation}
v_t=(h_v)_x, ~~~ w_t=(h_w)_y
\label{3H}
\end{equation}
which possess infinitely many hydrodynamic reductions. Here $h(v, w)$ 
is the Hamiltonian density. Notice that any system of the form 
(\ref{3H}) possesses one extra conservation law
$$
h_t=(h_v^2/2)_x+(h_w^2/2)_y.
$$
As mentioned in the introduction, Hamiltonian systems are related to 
the quadratic case of Sect. 3 by virtue of the Legendre transform. 
However, we find it instructive to treat the Hamiltonian case 
independently to better illustrate our method.

Looking for reductions in the form $v=v(R^1, ..., R^n)$, $w=w(R^1, 
..., R^n)$ where the Riemann invariants satisfy the equations
\begin{equation}
R^i_x=\lambda^i(R)\ R^i_t, ~~~~ R^i_y=\mu^i(R)\ R^i_t
\label{3R}
\end{equation}
and substituting into (\ref{3H}) one  arrives at the equations
$$
(1-\lambda^ih_{vv})\partial_iv=\lambda^ih_{vw}\partial_iw, ~~~
(1-\mu^ih_{ww})\partial_iw=\mu^ih_{vw}\partial_iv,
$$
(no summation!) so that $\lambda^i$ and $\mu^i$ satisfy the dispersion relation
$$
(1-\lambda^ih_{vv})(1-\mu^ih_{ww})=\lambda^i\mu^ih^2_{vw}.
$$
We require that the dispersion relation is nondegenerate (as a 
conic), that is, $h_{vw}\ne 0, \ h_{vw}^2-h_{vv}h_{ww}\ne 0$. Setting 
$\partial_iv=\varphi^i\partial_iw$, we obtain the following 
expressions for $\lambda^i$ and $\mu^i$ in terms of $\varphi^i$,
$$
\lambda^i=\frac{\varphi^i}{h_{vw}+\varphi^ih_{vv}}, ~~~~ 
\mu^i=\frac{1}{h_{ww}+\varphi^ih_{vw}},
$$
which define a rational parametrization of the dispersion relation. 
The compatibility conditions of the equations 
$\partial_iv=\varphi^i\partial_iw$ imply
\begin{equation}
\partial_i\partial_jw=\frac{\partial_j\varphi^i}{\varphi^j-\varphi^i}\partial_iw+\frac{\partial_i\varphi^j}{\varphi^i-\varphi^j}\partial_jw
\label{4w}
\end{equation}
while the commutativity equations (\ref{comm}) lead to the following 
complicated expressions for $\varphi^i$:
\begin{equation}
\begin{array}{c}
\partial_j\varphi^i=\frac{\varphi^i(h_{vw}+\varphi^jh_{vv})(h_{ww}+\varphi^ih_{vw})\partial_jw}{h_{vw}(h_{vw}^2-h_{vv}h_{ww})(\varphi^i-\varphi^j)}
\left[(h_{vvw}+\varphi^ih_{vvv})\varphi^j+h_{vww}+\varphi^ih_{vvw}\right]+ \\
\ \\
\frac{(h_{vw}+\varphi^ih_{vv})(h_{ww}+\varphi^jh_{vw})\partial_jw}{h_{vw}(h_{vw}^2-h_{vv}h_{ww})(\varphi^i-\varphi^j)}
\left[(h_{vww}+\varphi^ih_{vvw})\varphi^j+h_{www}+\varphi^ih_{vww}\right].
\end{array}
\label{4F}
\end{equation}
One can  see that   the compatibility condition 
$\partial_k\partial_j\varphi^i-\partial_j\partial_k\varphi^i=0$ is of 
the form
$P\partial_jw\partial_kw=0$ where $P$ is a rational expression in 
$\varphi^i, \varphi^j, \varphi^k$ whose coefficients depend on 
partial derivatives of the Hamiltonian density $h(v, w)$ up to fourth 
order. Requiring  that $P$ vanishes we obtain  the overdetermined 
system of fourth order PDEs for the density $h$:
\begin{equation}
\begin{array}{c}
h_{vw}(h_{vw}^2-h_{vv}h_{ww})h_{vvvv}=4h_{vw}h_{vvv}(h_{vw}h_{vvw}-h_{vv}h_{vww}) 
\\
+3h_{vv}h_{vw}h^2_{vvw}-2h_{vv}h_{ww}h_{vvv}h_{vvw}-h_{vw}h_{ww}h^2_{vvv}, \\
\ \\
h_{vw}(h_{vw}^2-h_{vv}h_{ww})h_{vvvw}=-h_{vw}h_{vvv}(h_{vv}h_{www} 
+h_{ww}h_{vvw})\\
+3h^2_{vw}h^2_{vvw}-2h_{vv}h_{ww}h_{vvv}h_{vww}+h^2_{vw}h_{vvv}h_{vww}, \\
\ \\
h_{vw}(h_{vw}^2-h_{vv}h_{ww})h_{vvww}=4h^2_{vw}h_{vvw}h_{vww} \\
-h_{vv}h_{vvw}(h_{vw}h_{www}+h_{ww}h_{vww})-h_{ww}h_{vvv}(h_{vw}h_{vww}+h_{vv}h_{www}), 
\\
\ \\
h_{vw}(h_{vw}^2-h_{vv}h_{ww})h_{vwww}=-h_{vw}h_{www}(h_{ww}h_{vvv}+h_{vv}h_{vww}) 
\\
+3h^2_{vw}h^2_{vww}-2h_{vv}h_{ww}h_{www}h_{vvw}+h^2_{vw}h_{www}h_{vvw}, \\
\ \\
h_{vw}(h_{vw}^2-h_{vv}h_{ww})h_{wwww}=4h_{vw}h_{www}(h_{vw}h_{vww}-h_{ww}h_{vvw}) 
\\
+3h_{ww}h_{vw}h^2_{vww}-2h_{vv}h_{ww}h_{www}h_{vww}-h_{vw}h_{vv}h^2_{www}. \\
\end{array}
\label{4int}
\end{equation}
It was verified   that this system is in involution and its  solution 
space is  10-dimensional. As mentioned in the introduction, the 
Legendre transform identifies the systems
(\ref{4int}) and (\ref{f}).

\section{Concluding remarks}

We have demonstrated that the existence of  
 hydrodynamic reductions describing nonlinear interactions 
of $n\geq 3$ planar simple  waves can be viewed as  the effective integrability 
criterion. The most natural problems arising in this context are the 
following:

\noindent 1. Classify multicomponent  $(2+1)$-dimensional integrable 
quasilinear systems. The main difference from the  two-component case 
is that the dispersion relation
(\ref{dispersion}) will no longer define a rational curve.

\noindent 2. The recent publication \cite{Fer2} suggests that the 
method of hydrodynamic reductions carries over to $3+1$ dimensions. 
It would be extremely interesting to obtain further examples 
(classification results) of dispersionless integrable systems in many 
dimensions.

\noindent 3. Nonlinear interactions of $n$  simple waves can be 
viewed as a natural dispersionless analogue of `$n$-gap' solutions. 
It would be desirable to obtain an alternative description of these 
solutions as the `stationary points' of the appropriate `higher 
symmetries'.

We hope to address these questions elsewhere.

\section*{Acknowledgements}

EVF is grateful to  A. Fokas for drawing his attention to the problem 
of integrability of multi-dimensional quasilinear systems. We also 
thank C. Dafermos, M. Dunajski, J. Gibbons, B. Keyfitz,  M. Pavlov, D. Serre and 
C. Trivisa for their interest and clarifying discussions. This research was initiated 
when both authors were attending the research semester on Hyperbolic 
Conservation Laws at the Isaak Newton Institute in the spring of 
2003. It is a great pleasure to thank P. LeFloch for the invitation 
to participate in this program.

\end{document}